\documentclass{emulateapj}

\slugcomment{Draft version \today}

\shorttitle{Iron dust in globular clusters}
\shortauthors{McDonald et al.}

\begin{document}

\title{Rusty old stars: a source of the missing interstellar iron?}

\author{I.~McDonald\altaffilmark{1}, G.~C.~Sloan\altaffilmark{2}, A.~A.~Zijlstra\altaffilmark{1}, N.~Matsunaga\altaffilmark{3}, M.~Matsuura\altaffilmark{4,5}, K.~E.~Kraemer\altaffilmark{6}, J.~Bernard-Salas\altaffilmark{2}, A.~J.~Markwick\altaffilmark{1}}
\altaffiltext{1}{Jodrell Bank Centre for Astrophysics, Alan Turing Building, Manchester, M13 9PL, UK; iain.mcdonald-2@jb.man.ac.uk, albert.zijlstra@manchester.ac.uk, andrew.markwick@manchester.ac.uk}
\altaffiltext{2}{Cornell University, Astronomy Department, Ithaca, NY 14853-6801; sloan@isc.astro.cornell.edu}
\altaffiltext{3}{Institute of Astronomy, University of Tokyo, 2-21-1 Osawa, Mitaka, Tokyo 181-0015, Japan; matsunaga@ioa.s.u-tokyo.ac.jp}
\altaffiltext{4}{UCL-Institute of Origins, Astrophysics Group, Department of Physics and Astronomy, University College London, Gower Street, London WC1E 6BT, United Kingdom} 
\altaffiltext{5}{UCL-Institute of Origins, Mullard Space Science Laboratory, University College London, Holmbury St. Mary, Dorking, Surrey RH5 6NT, United Kingdom; mikako@star.ucl.ac.uk}
\altaffiltext{6}{Air Force Research Laboratory, Space Vehicles Directorate, Hanscom AFB, MA 01731}

\begin{abstract}
Iron, the Universe's most abundant refractory element, is highly depleted in both circumstellar and interstellar environments, meaning it exists in solid form. The nature of this solid is unknown. In this Letter, we provide evidence that metallic iron grains are present around oxygen-rich AGB stars, where it is observationally manifest as a featureless mid-infrared excess. This identification is made using \emph{Spitzer Space Telescope} observations of evolved globular cluster stars, where iron dust production appears ubiquitous and in some cases can be modelled as the only observed dust product. In this context, FeO is examined as the likely carrier for the 20-$\mu$m feature observed in some of these stars. Metallic iron appears to be an important part of the dust condensation sequence at low metallicity, and subsequently plays an influential r\^{o}le in the interstellar medium. We explore the stellar metallicities and luminosities at which iron formation is observed, and how the presence of iron affects the outflow and its chemistry. The conditions under which iron can provide sufficient opacity to drive a wind remain unclear.
\end{abstract}



\section{Introduction}

Iron is the sixth most abundant element in the Universe, and the most abundant
refractory element. It is observed to be highly depleted in both interstellar
and circumstellar environments \citep{SB79,MvWLE05,DIRMV09}, and must
therefore predominantly exist in an unknown solid form. Iron may be
incorporated in other dust grains, primarily silicates, though these are
usually iron-poor \citep{GS99,KdKW+02}. Alternatively, iron may form a
metallic condensate \citep{KdKW+02,VvdZH+09}.


Dust grains form around evolved stars by condensation from the gas phase,
either directly, or onto molecular `seeds'. Individual dust species can
usually be identified by distinct infrared emission bands. In oxygen-rich
environments, amorphous or crystalline forms of silicates, spinels and
corundum are observed to form, whilst carbon-rich environments give rise to
amorphous carbon, graphite and SiC. Metallic iron grains, however, produce
featureless infrared emission which can be difficult to differentiate from
other sources, particularly amorphous carbon dust. Iron has hitherto only been
inferred as a likely component of oxygen-rich dusty winds
\citep{KdKW+02,VvdZH+09}, but never positively identified.

Previous observational studies \citep{MvLD+09,BMvL+09} of giant stars in globular clusters have found a featureless contribution in addition to the flux emanating from the star's photosphere, which has been attributed to an unidentified circumstellar dust species. We herein show that this dust species is metallic iron.


\section{Evidence for dust emission}
\label{AnalSect}

\subsection{Observations}

Our sample consists of 35 highly-evolved, (strongly-)pulsating stars
in globular clusters \citep{SMM+10} (Paper I), observed using the \emph{Spitzer Space Telescope}
Infrared Spectrograph (IRS). These long-period variable
stars lie near or above the red giant branch tips of 19 Galactic globular
clusters, which range in metallicity from [Fe/H] = --1.62 to --0.10. Observed
with \emph{Spitzer's} low-resolution modules, the spectra cover $\lambda = 5.2
- 38$ $\mu$m with a resolving power of $R \sim 60 - 120$. 

One carbon-rich star (Lyng\aa\ 7 V1) is discounted, leaving a sample of 34 stars. Of these, silicate features at 10 and 20 $\mu$m are seen in 24 objects, implying oxygen-rich chemistry. More pertinently, the remaining ten stars either have very weak silicate features or appear to be `naked'.

$JHKL$-band light curves are also available (Paper I) for 25 of our 34 objects, which allows us to construct the photometric fluxes of these objects at the pulsation phase of their IRS spectrum.

\begin{figure}
\includegraphics[width=0.35\textwidth,angle=-90]{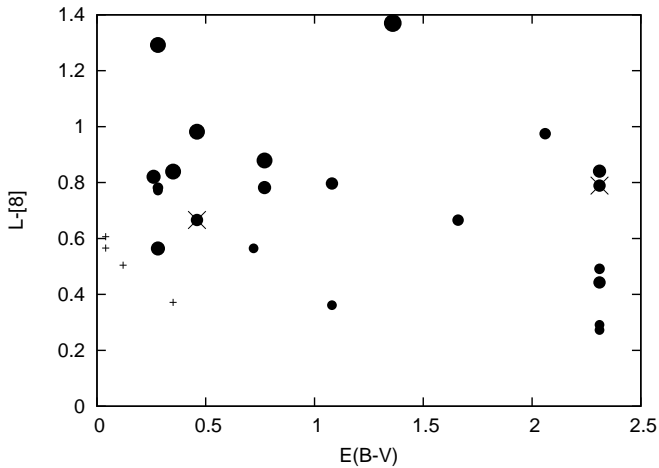}
\caption{Extinction-corrected $L$--[8] colours in our sources. Plus signs show stars with pure iron dust, filled circles show stars also hosting silicate dust (symbol size is proportional to strength of emission). Crosses show stars exhibiting a strong 20-$\mu$m feature. Colour does not vary significantly with $E(B-V)$, showing that this is not merely a reddening effect.}
\label{Excess2Fig}
\end{figure}

Figure \ref{Excess2Fig} shows the $L$--[8] colour for these objects, where the
8-$\mu$m photometry was derived by convolving the IRS spectrum with the
\emph{Spitzer} IRAC 8-$\mu$m filter. A truly naked photosphere should have
$L-[8] \approx 0$, once interstellar reddening has been taken into
account\footnote{We here assume $E$(L--[8]) = $0.017E(B-V)$, after
  \citet{Whittet92} and \citet{FPM+07}.} (c.f.\ \citealt{BMvL+08}). All 25 stars, however, have a positive colour. This implies a flux excess at 8 $\mu$m
which cannot be caused by commonly-assumed dust species (silicates, alumina)
under normal conditions of temperature and grain size. This excess flux at long wavelengths is confirmed by comparison to {\sc marcs} photospheric model spectra \citep{GBEN75,GEE+08,MvLD+09} as shown below.

From our sample, four example stars have been selected for detailed modelling:
NGC 5927 V3, a highly-evolved AGB star with a strong, narrow, 10-$\mu$m
silicate feature; NGC 6352 V5, a less luminous giant star with unusual dust
features; and NGC 362 V2 and NGC 5139 V42, `naked' RGB-tip or AGB stars
showing optical TiO bands (indicating they are oxygen-rich) and infrared dust
excesses, but no silicate features (\citealt{SSK99}; \citealt{MvLD+09};
\citealt{BMvL+09}).

\subsection{Fitting the observations}

\begin{figure*}
\includegraphics[width=0.35\textwidth,angle=-90]{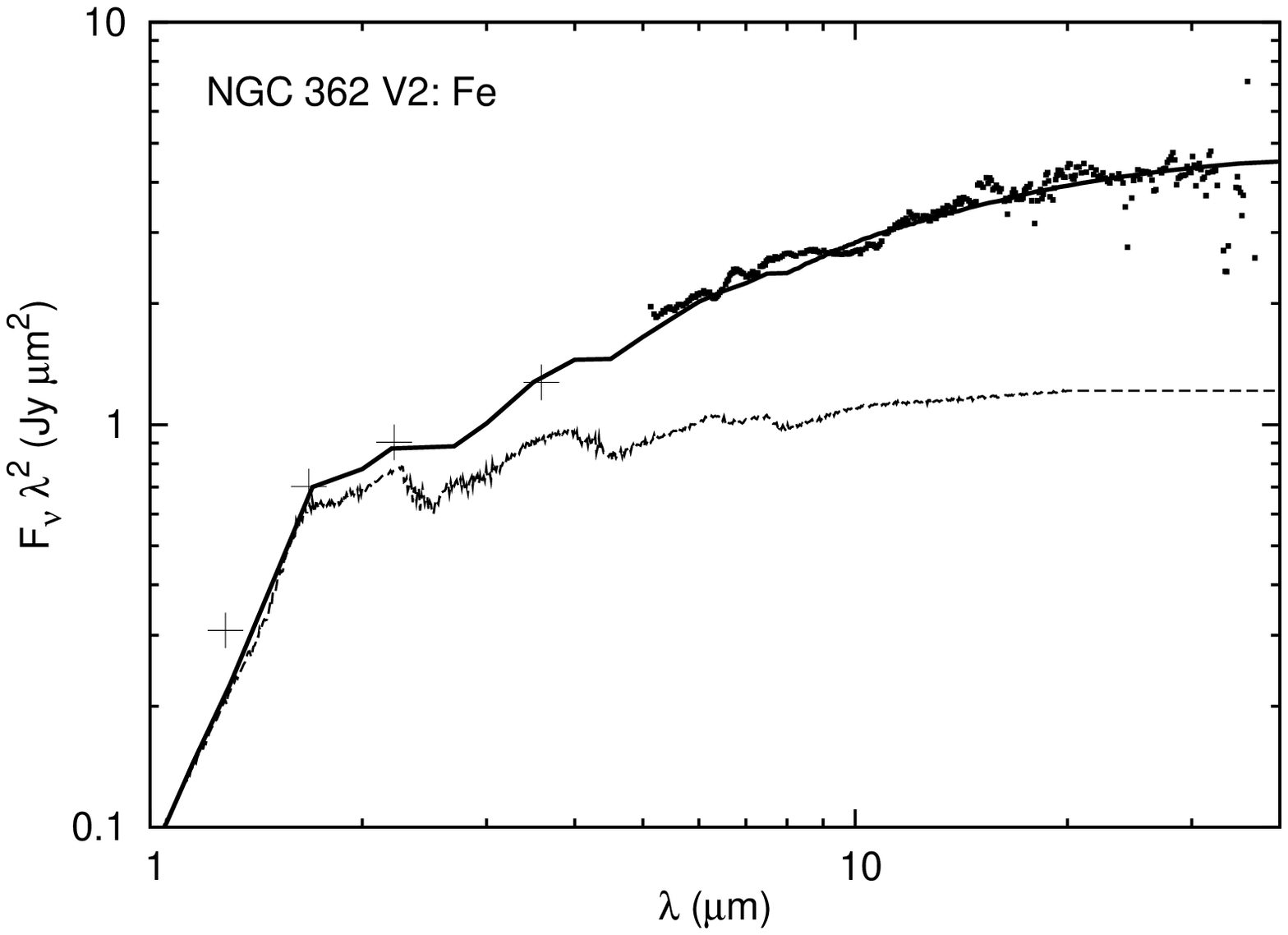}\includegraphics[width=0.35\textwidth,angle=-90]{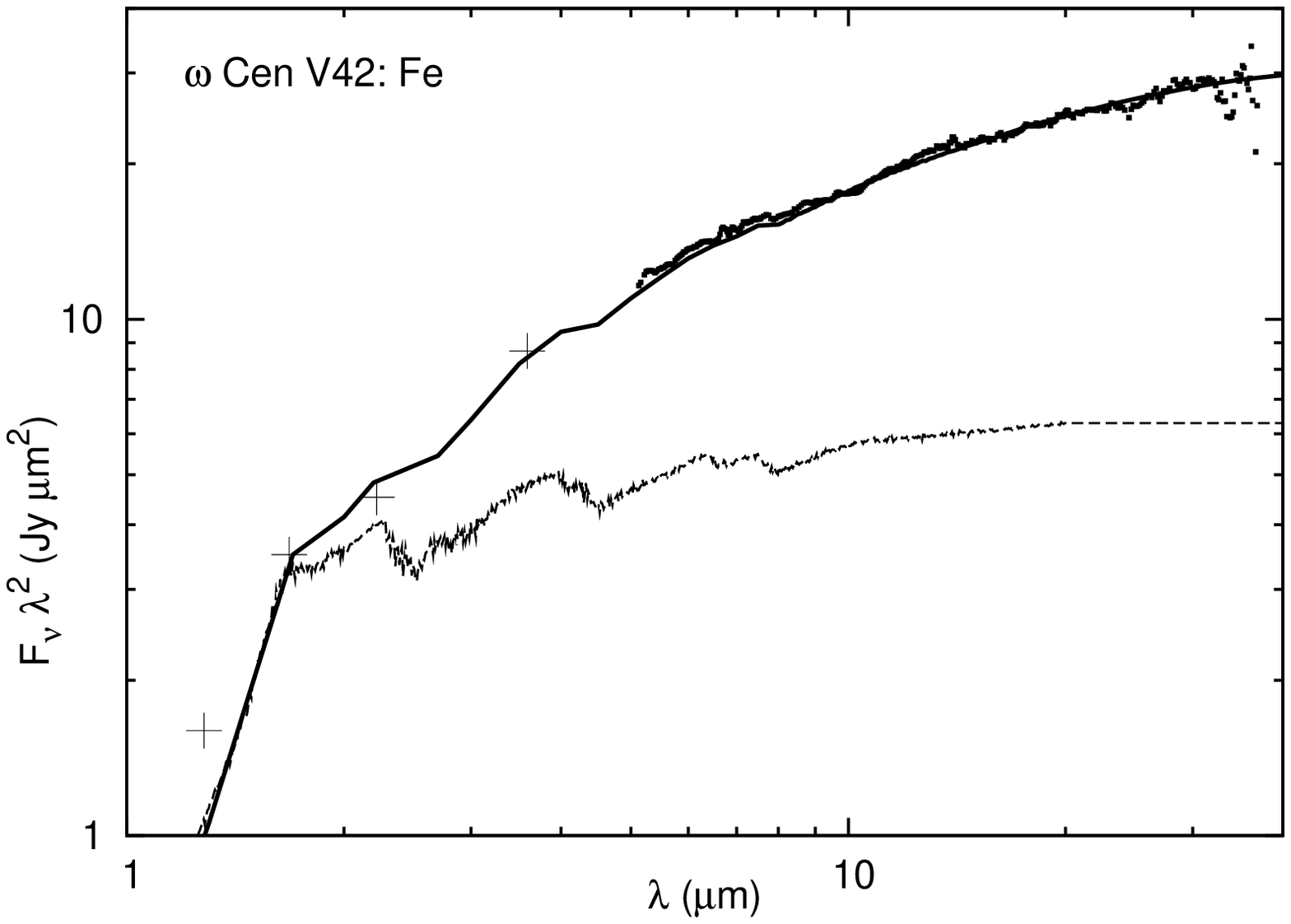}
\includegraphics[width=0.35\textwidth,angle=-90]{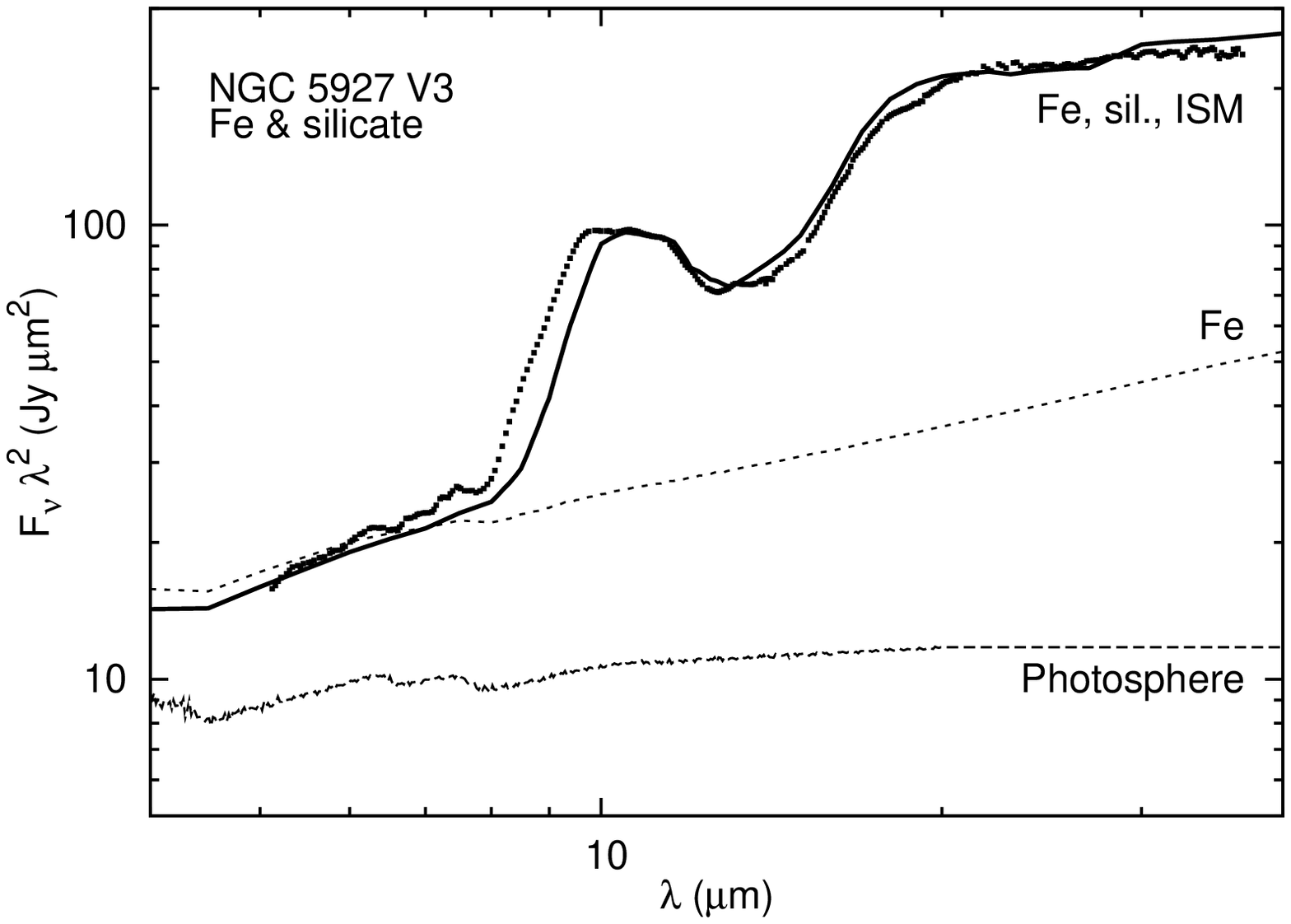}\includegraphics[width=0.35\textwidth,angle=-90]{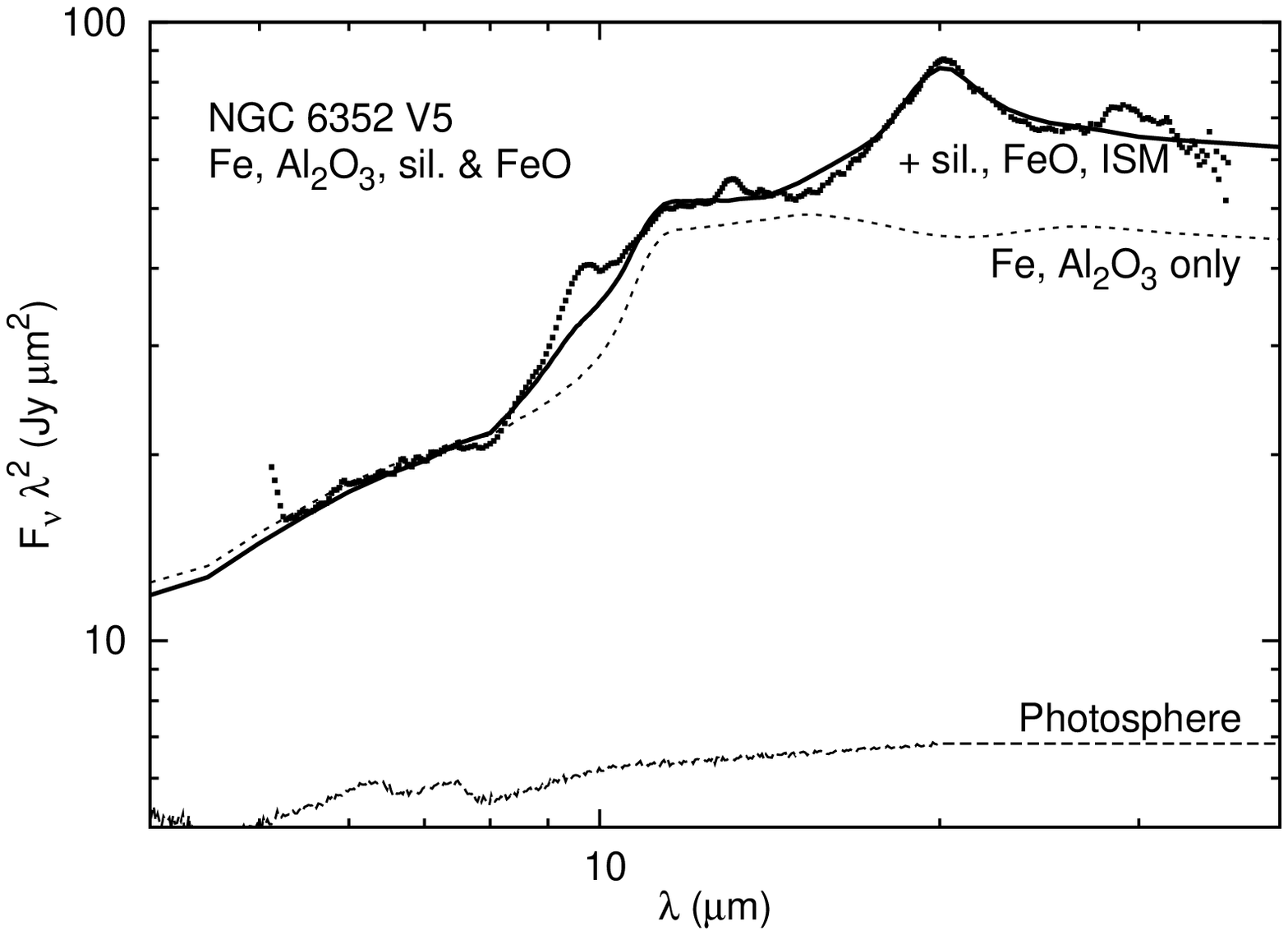}
\caption{Modelled dust contributions (solid lines) to \protect\emph{Spitzer} IRS spectra (black dots) and $JHKL$ photometry (black pluses) which has been corrected to the pulsation phases of the IRS spectra. The dashed line is a {\sc marcs} model denoting the photospheric contribution. The bottom two panels include the extra indicated contributions from Al$_2$O$_3$, silicates and FeO (dotted lines).}
\label{ExampleFig}
\end{figure*}

\begin{deluxetable*}{lcccccc}[ht!]
\tablecolumns{7} \tablewidth{0pc} \tabletypesize{\footnotesize}
\setlength{\tabcolsep}{0.0in} \renewcommand{\arraystretch}{1.0}
\tablewidth{0pt} \tablecaption{Parameters of {\sc dusty} models presented in Figure \ref{ExampleFig}. \label{ExampleTable}}
\tablehead{\colhead{Parameter} & \colhead{NGC 362 V2} & \colhead{$\omega$ Cen V42} & \multicolumn{2}{c}{NGC 5927 V3} & \multicolumn{2}{c}{NGC 6352 V5} \\ }
\startdata 

Assumed luminosity (L$_\odot$)			& 1826	& 1862	& \multicolumn{2}{c}{2000}	& \multicolumn{2}{c}{2000} \\
Assumed [Fe/H]					& --1.16& --1.62& \multicolumn{2}{c}{--0.37} 	& \multicolumn{2}{c}{--0.70} \\
Assumed $E(B-V)$				& 0.05	& 0.10	& \multicolumn{2}{c}{0.45} 	& \multicolumn{2}{c}{0.21} \\
References					& 1,3	& 2	& \multicolumn{2}{c}{4}		& \multicolumn{2}{c}{4} \\

\noalign{\vskip 0.5ex}
\hline
\noalign{\vskip 0.5ex}
\ 						&	&	&\multicolumn{2}{c}{$\overbrace{\qquad\qquad\qquad}$}	&\multicolumn{2}{c}{$\overbrace{\qquad\qquad\qquad\qquad\qquad}$}\\
Wind component 					& 	& 	& A	& B	& A	& B\\
\noalign{\vskip 0.5ex}
\hline
\noalign{\vskip 0.5ex}
Composition 					& 100\%	& 100\%	& 100\%	& 100\%	& 96\% Fe	& 70\% Silicate\\
						& iron	& iron	& iron	& silicate	& 4\% Al$_2$O$_3$	& 30\% FeO\\

\noalign{\vskip 1ex}
$T_{\rm inner\ dust\ envelope\ edge}$ (K)	& 800	& 1000	& 1100	& 400	& 1000	& 450	\\
$T_{\rm outer\ dust\ envelope\ edge}$ (K)	& 194	& 173	&\nodata&\nodata&\nodata&\nodata\\
$r_{\rm outer}/r_{\rm inner}$			& 50	& 120	& 1000	& 1000	& 100	& 100	\\
Optical depth at 0.55 $\mu$m			& 0.28	& 0.70	& 0.55	& 0.45	& 1.05	& 0.12	\\
\enddata 
\tablenotetext{}{{\bf References:} (1) \protect\cite{BMvL+09}; (2) \protect\cite{MvLD+09}; (3) $E(B-V)$ from \protect\cite{Harris96}; (4) luminosity estimated, metallicity from \protect\cite{Harris96}.}
\end{deluxetable*}

For a dust-free comparison, we use a stellar atmosphere model created using
the {\sc marcs} code \citep{GBEN75,GEE+08} at 3500 K, [Fe/H] =
--1.0, [$\alpha$/Fe] = +0.3 (see \citealt{MvLD+09} for details of these
models). Being very similar objects near the giant branch tip, we expect the
effective surface temperature \emph{below the dust-producing zone} to be 3750
$\pm \sim$250 K. We use the near-infrared flux to fit the photospheric
continuum. A low-temperature, metal-rich model therefore represents an
intentional bias toward a model with a more-luminous infrared spectrum
compared to its near-infrared flux. This reduces the calculated excess to rule
out disparity between the modelled and real photospheric spectra.

Figure \ref{ExampleFig} shows the observed spectra, alongside fits to the
stellar wind contribution using {\sc dusty} \citep{NIE99}. Using the {\sc
  marcs} model atmosphere as an input, we fit a stellar wind model using
metallic iron grains \citep{OBA+88}. For each model we assume a
radiatively-driven wind; and a standard Mathis-Rumpl-Nordsieck grain size
distribution \citep{MRN77}, given by $n(a) = a^{-q}$, $q = 3.5$ for $a = 5 -
250$ nm.

We correct for interstellar reddening using the absorption profiles from \citet{McClure09}. We further assume that $A_{\rm V} = 3.2 E(B-V)$ and that $A_{\rm V} / A_{\rm K} = 7.75$. $E(B-V)$ values for NGC 362, 5927 and 6352 are from \citet{Harris96}, $E(B-V) = 0.10$ is assumed for $\omega$ Cen, based on \citet{Harris96} and \citet{MvLD+09}. The correction applied is thus quite small and negligibly different from that applied in Paper I.

NGC 362 V2 and $\omega$ Cen V42 are best fit with a dust shell truncated to an
outer radius ($r_{\rm outer}$) of $\sim$100 times the inner radius ($r_{\rm
  inner} =$ the dust formation radius). (Normally, fits are extended to
$r_{\rm outer} / r_{\rm inner} \gtrsim 1000$.) This suppresses the flux at
$\lambda > 30 \mu$m. A similar effect can also be achieved by using a grain
distribution with fewer large grains: both effects may be in play (we only
model a truncated shell here). The strength, and indeed existence of, this
suppression depends crucially on the subtraction of the infrared background
(caused by Galactic interstellar medium and other cluster objects). While
every effort has been made to make sure this has been done correctly, we
cannot be conclusively sure that such a suppression and thus truncation exist.

In NGC 5927 V3 and NGC 6352 V5, we also observe other dust species, which emit
at $\lambda > 8 \mu$m. Identified species include silicates (magnesium-rich
olivines and pyroxenes) at 10 and 20 $\mu$m and amorphous alumina
(Al$_2$O$_3$) at 11--15 $\mu$m (optical constants from \citealt{DL84};
\citealt{BDH+97}). Related crystalline minerals with a smaller degree of
amorphisation are likely responsible for the sharp 9.6- and 13.1-$\mu$m
features seen in NGC 6352 V5 (\citealt{SKGP03}). Note that the fit to NGC 5927
V3 produces a fairly good match to the 10-$\mu$m feature, but at a wavelength
which is too red. This can be caused by a number of factors, including grain
size and porosity \citep{VH08}. The 20-$\mu$m feature has an unknown carrier: a promising candidate is magnesio-w\"{u}stite (Mg$_x$Fe$_{1-x}$O; see, e.g., \citealt{PKM+02}; \citealt{LPH+06}; Paper I). This provides a good fit to the feature in NGC 6352
V5. The comparatively-red peak wavelength seen in NGC 6352 V5 further suggests
that $x \approx 0$: i.e.\ that it is near-pure FeO \citep{HBMD95}. We return
to this feature later.

We could not fit a single-temperature dust model to these two stars, implying
that the dust species are not in thermal contact with each other. Iron and
alumina were fit at temperatures of $\sim$1000 K. In NGC 6352 V5, silicates
and FeO were fit at $\sim$450 K; in NGC 5927 V3, the strong 10-$\mu$m peak
suggests mainly glassy silicates, which we fit at $\sim$400 K.

Table \ref{ExampleTable} lists our fitted parameters, including dust temperatures at the outer and inner edges of the dust envelope ($T_{\rm outer}$, $T_{\rm inner}$) and the size of the envelope used ($r_{\rm outer}/r_{\rm inner}$, $r_{\rm inner} \sim 2$ R$_\ast$). We do not include mass-loss rates, as explained at the end of this section.

\subsection{Alternative sources of excess}

\begin{figure*}
\includegraphics[width=0.35\textwidth,angle=-90]{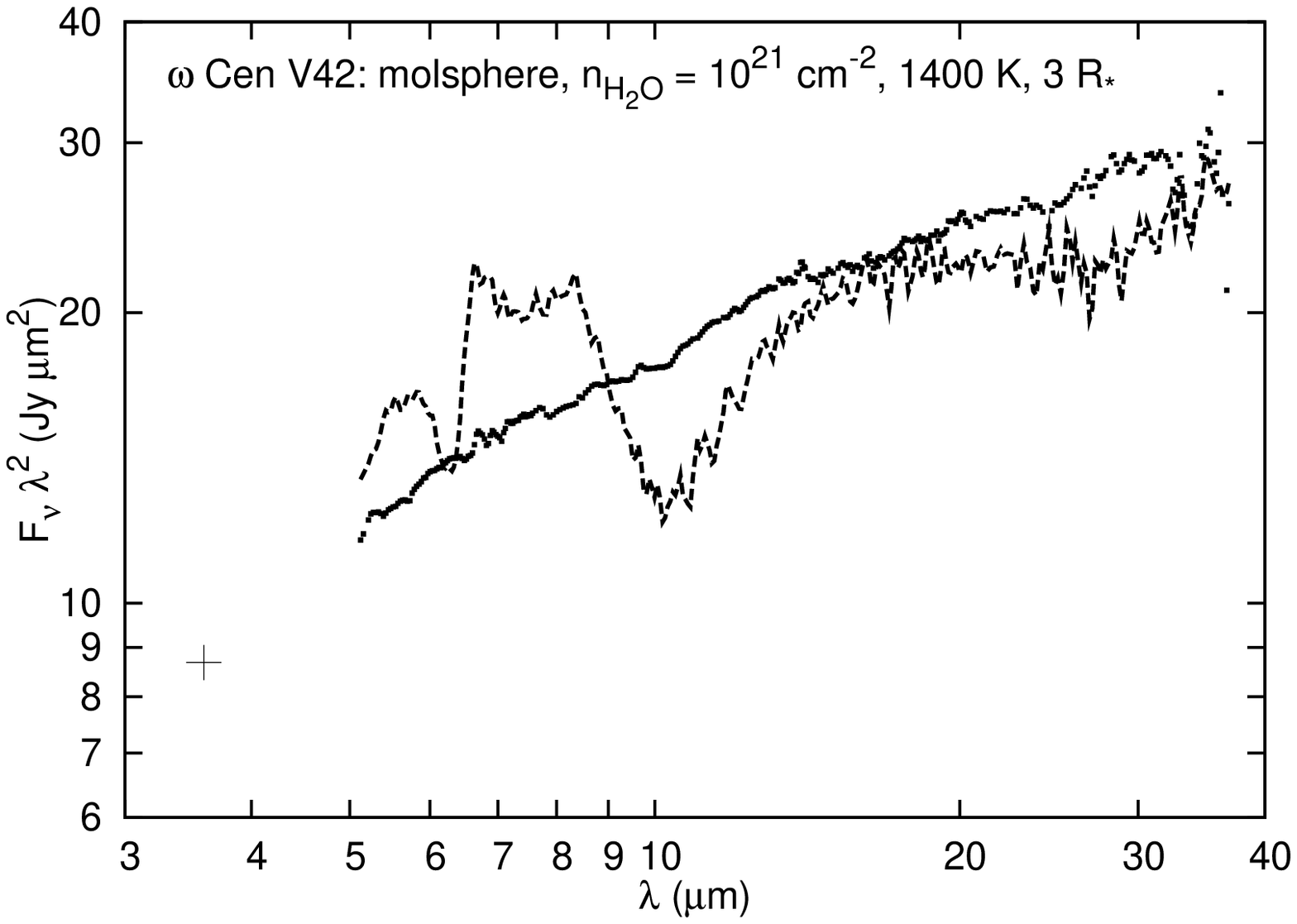}
\includegraphics[width=0.35\textwidth,angle=-90]{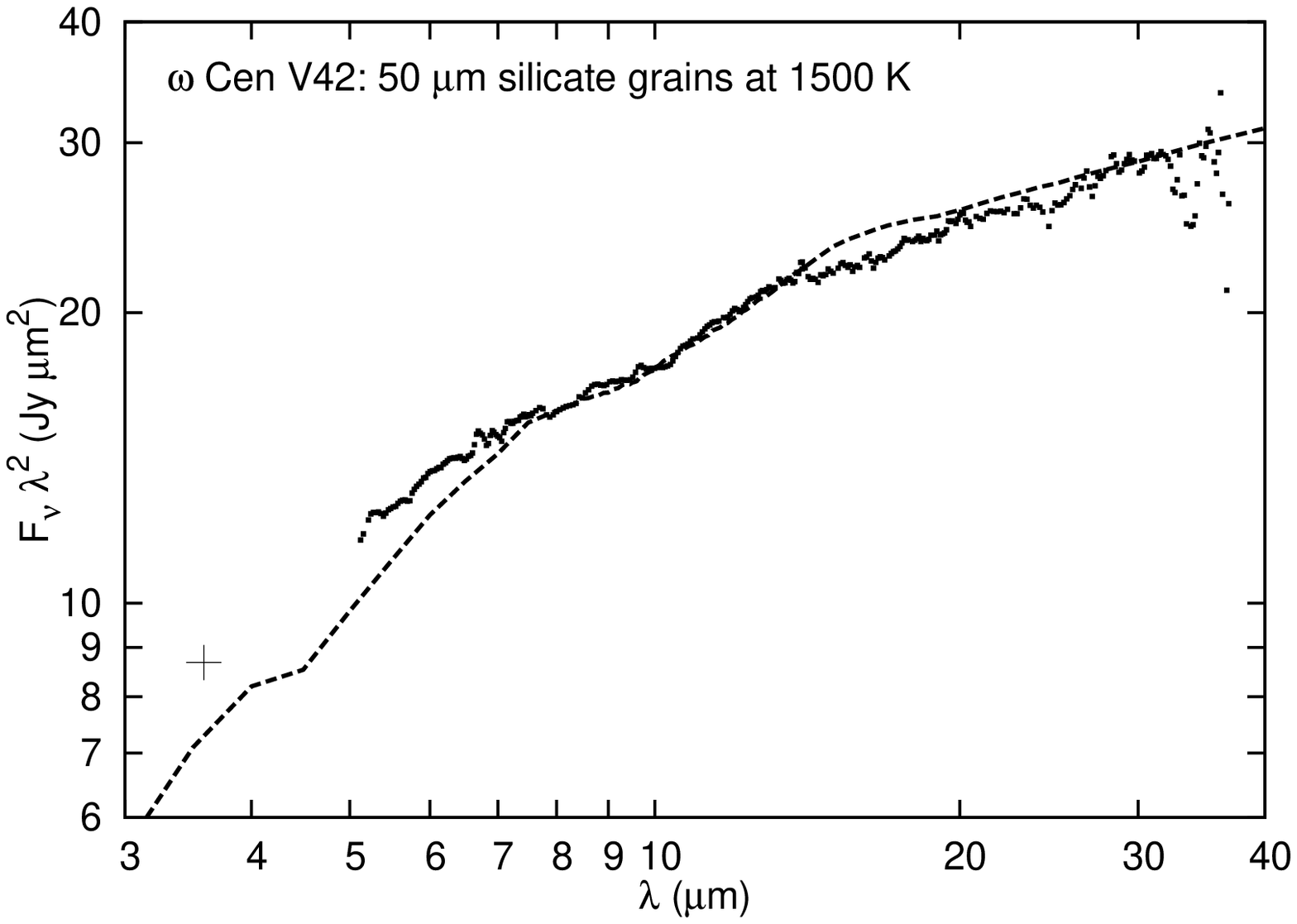}
\includegraphics[width=0.35\textwidth,angle=-90]{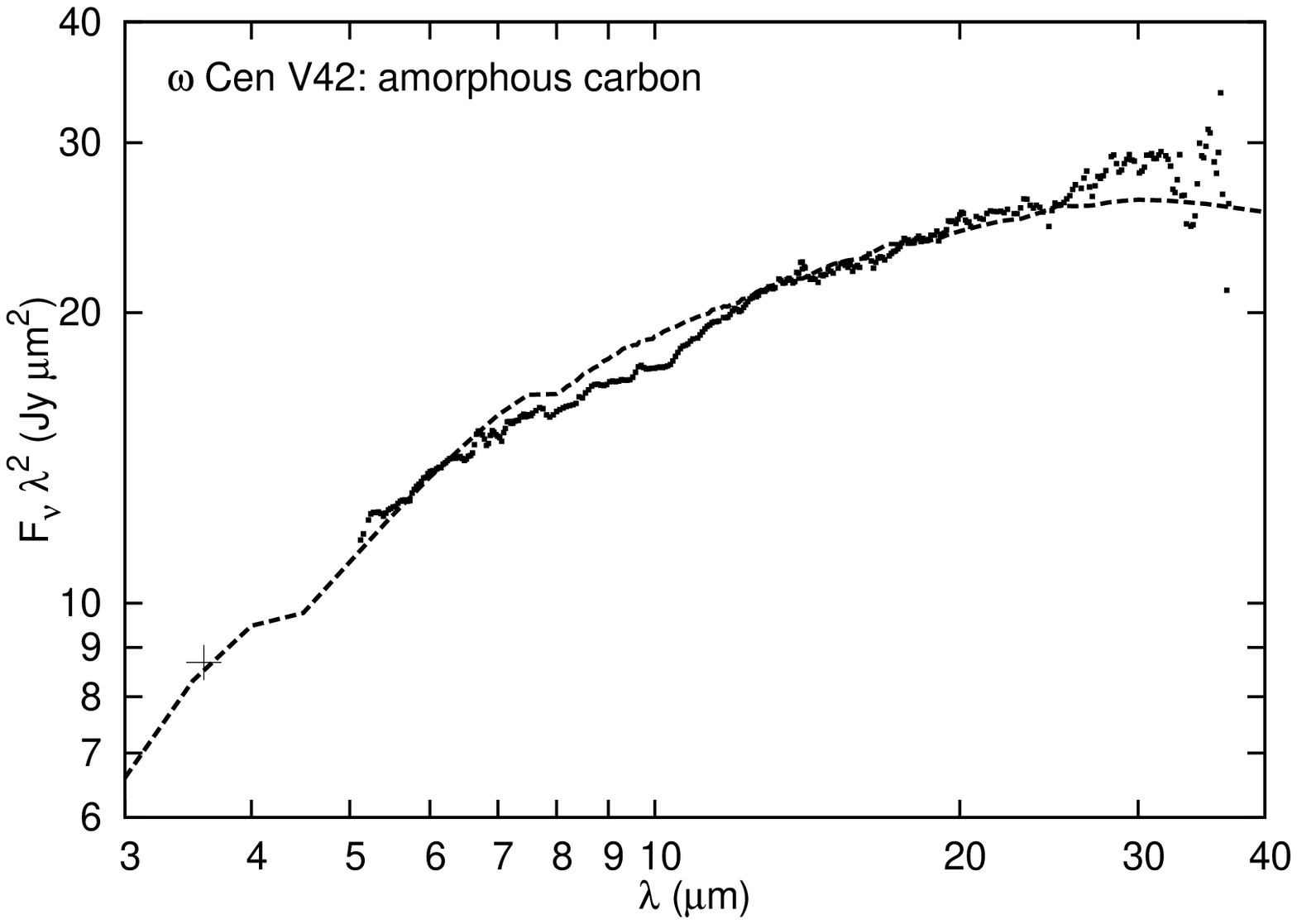}
\includegraphics[width=0.35\textwidth,angle=-90]{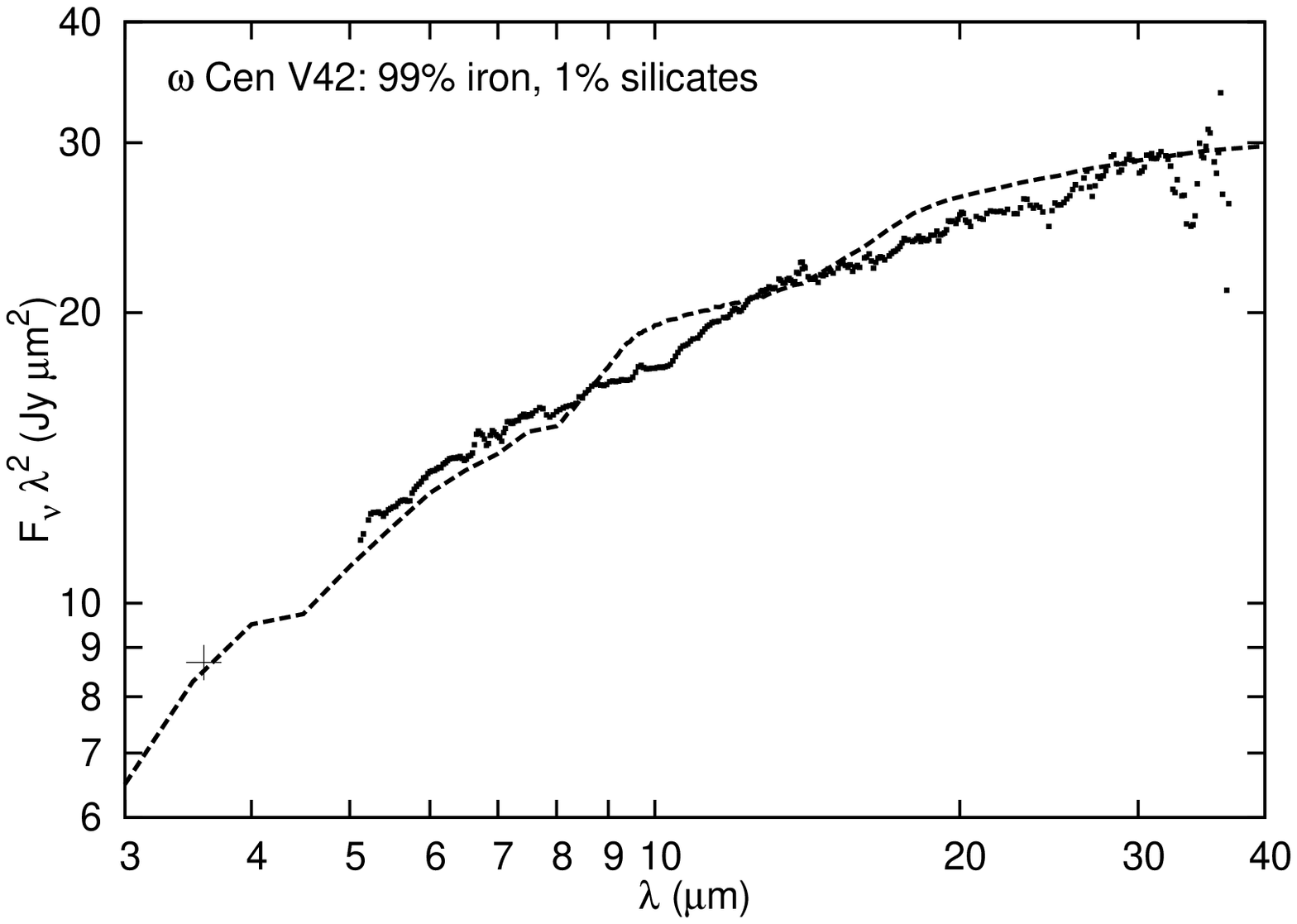}
\caption{Fits to the \emph{Spitzer} IRS spectrum of $\omega$ Cen V42 using alternative emission sources. In each panel, the following is shown: dots --- IRS spectrum; plus sign --- phase-corrected $L$-band photometry; dotted line --- fit for indicated grain composition or emission mechanism.}
\label{ModelsFig}
\end{figure*}

Featureless infrared excess can be produced by a variety of circumstellar
material. This can take the form of free-free emission, emission from shells
of molecular gas (a `molsphere'; e.g.\ \citealt{Tsuji00}), extremely-large silicate
grains, or amorphous carbon grains. Figure \ref{ModelsFig} shows modelled fits
for these possible emission sources, modelled in an identical way to the iron
fits listed above.

Ionisation in a chromosphere or shock fronts propagating through the outer atmosphere can give rise to free-free emission, as observed by radio observations toward Galactic Miras \citep{RM97}. The optical depth of free-free emission in a region of thickness $r$ can be calculated as:
\begin{equation}
\tau_\nu = \int_r 3.28 \times 10^{-7} \left(\frac{T_{\rm e}}{10^4 {\rm K}}\right)^{-1.35} \!\! \nu_{\rm GHz}^{-2.1} \left(\frac{n_{\rm e}}{\rm cm^{-3}}\right)^2 \frac{\delta r}{\rm pc}
\end{equation}
where the electron density, $n_{\rm e}$, is recoverable from the electron temperature, $T_{\rm e}$, via the Saha equation. At low frequencies, where $\tau_\nu \gg 1$, the emission scales as $F_\nu \propto \nu^{-2}$ for an isothermal, isobaric region; and $F_\nu \propto \nu^{-0.6}$ for an isothermal wind expanding with constant velocity. For $\omega$ Cen V42, we find our spectrum would require $\tau_\nu = 1$ at 3--4 $\mu$m, and that $F_\nu \propto \nu^{-1.6}$ for $\tau_\nu \gg 1$. This more-closely approximates an isobaric density profile, which would be difficult to achieve in an expanding wind. Very similar conditions would have to exist in all 34 of our stars, as the shape of their spectra are very similar.

An isobaric chromosphere producing a spectrum with this optical depth cannot exist except in the following case: the wind must be mostly ionised ($\gtrsim 6000$ K) {\it and} either the chromosphere extends to an unphysical radius ($\gtrsim 70$ R$_{\odot}$) and/or the mass-loss rate becomes unphysical ($\gtrsim 10^{-5.5}$ M$_\odot$ yr$^{-1}$) for a thermally-expanding wind. Departing from an isobaric state makes conditions even more prohibitive. While chromospheres may be present in these stars, this infrared excess is not seen in chromospherically-active stars that are less evolved \citep{MvL07,BMvL+09}. In V42, circumstellar ionisation appears dominated by pulsation shocks. \citet{RM97} show that, for Galactic Miras, the temperature and density increases provided by shocks in the outer atmosphere are insufficient to produce an ionised layer that would be optically thick in the infrared. We thus rule out free-free emission as a dominant contributor to our spectra.

The molsphere creates a dense `forest' of molecular lines. Water lines are the most important, and cause emission throughout the spectrum, but mainly in the 6--8-$\mu$m and $>20 \mu$m regions. This emission is seen in some of our example stars (Paper I). However, the water bands in our spectra are very weak compared to stars with naked molspheres (e.g.\ $\beta$ Peg --- \citealt{TOAY97}). Gas-phase model emission spectra (Figure \ref{ModelsFig}), created from the line list of \citet{PS97} using SpectraFactory\footnote{http://spectrafactory.net} \citep{CvMM10}, fail to reproduce the flatness of the spectrum for any reasonable combination of temperature and column density. Figure \ref{ModelsFig} shows the highest column density we consider feasible, and even this contains $\sim 10^{-7}$ M$_\odot$ of water or $\sim$0.01\% of the star's oxygen. Furthermore, most molsphere host stars still require an extra opacity source (such as iron grains) providing flux at 6--8$\mu$m \citep{VvdZH+09}. Infrared molecular emission should therefore not make a significant contribution to our spectra.

Of the commonly-considered dust species only metallic iron and amC produce a featureless infrared continuum that matches the observed `naked' spectra. Other dust species (e.g.\ graphite) produce either strong spectral features or inflections which do not fit observed spectra. Spectral features can be damped by unusual conditions, notably exceptionally large grains \citep{Hoefner08}: these cannot simultaneously reproduce both the excess at $\lambda < 8 \mu$m and $\lambda > 20 \mu$m at the observed levels (Figure \ref{ModelsFig}) and are not expected in metal-poor stars where grain growth may be hampered by a lack of constituent elements.

Observationally, amC can also provide an acceptable fit to our example stars
over most of the spectrum. The remaining discrepancies may be fixable by
fine-tuning the wind parameters (grain size, density distributions,
temperatures, background subtraction, etc.).  Detection of amC as dominant
dust species in an oxygen-rich environment would be unexpected. In
giant stars, carbon and oxygen are bound into CO near the stellar surface,
locking away the least abundant element. The remaining majority element,
either carbon or oxygen, dominates the dust chemistry. CO may be dissociated in atmospheric shock waves or by chromospheric UV irradiation \citep{BGH+92}. This can lead to the production of small amonuts of other carbon-rich molecules: mainly CO$_2$ (observed in our stars: see Paper I), but also HCN and CS \citep{DCW99}. Small quantities of carbon-rich dust may then form \citep{HA07}, though oxygen-rich dust production is still expected to be dominant. In our sample, however, the featureless component often appears to be the dominant species, and no evidence for oxygen-rich dust is found.

Having ruled out the alternatives, we conclude that the source of the dust
excess must be iron grains, and that these must dominate the condensed
material in some stars. If we na\"{i}vely assume identical populations of
spherical grains, the fraction of warm silicates in the `naked' stars being
$\lesssim$ 1\% of the iron content (Figure \ref{ModelsFig}; the same would be
true at a $\lesssim$ 3\% level for an amC component). The silicate fraction
may increase if iron grains are made smaller and oblate, thus presenting a
higher cross-section per unit mass: extreme needle-like grains are required if
silicates are to become the dominant dust species by mass. As we do not know the grain shape, it becomes difficult to estimate a mass-loss rate for an iron wind. Spherical grains produce unfeasibly low expansion velocities of order $\sim$m
s$^{-1}$ for a pure iron wind; expansion velocities for NGC 5927 V3 and NGC 6352 V5 were not calculated as the model assumes a zero initial velocity for both temperature
components, which will not be the case if they form at different radii. The
mass-loss rates and expansion velocities can also be significantly modified by
altering, e.g., the grain size or density (single crystal
vs. conglomerate), but reasonable changes still fail to produce feasible
values, except for extremely elongated grains.

This implies that iron alone may not be able to drive a dusty wind in these stars. Other factors may also accelerate the wind. These could include small amounts of amorphous carbon \citep{HA07}; energy deposition either from shocks created by stellar pulsations (\citealt{Bowen88}), or magnetic reconnection (\citealt{PH89}); increased opacity from circumstellar molecular gas (\citealt{EBJ89}); stellar pulsation, either by direct acceleration or dissipation of shocks; line-driving of H {\sc i} or Ca {\sc ii} atoms, or H$_2$O molecules (e.g.\ \citealt{EBJ89}; \citealt{Bowen89}). The driving mechanism is thus likely to be complex, with radiation pressure on metallic iron being only one of several accelerants which drive matter from the star. Obtaining an accurate radial profile of the wind's outflow velocity will be crucial in determining which factors dominate this acceleration.


\section{Discussion}

\subsection{The fate of iron and iron oxide grains}

Efficient production of metallic iron by oxygen-rich AGB stars would explain several iron depletions. Gas-phase depletion of iron extends to the interstellar medium \citep{SB79}. Iron grains contribute to infrared emission from this medium \citep{CL88} and iron is seen to be destroyed in shocks \citep{LBSB+08}. Iron is also depleted by $>90$\%\ in the gas phase of planetary nebulae \citep{DIRMV09}. Direct evidence for iron condensation in AGB winds comes from depletion patterns in binary post-AGB stars. In these stars, the wind is captured in a circumbinary disk, from where gas re-accretes onto the star. These stars show large depletions of refractory elements, with iron being among the most depleted \citep{MvWLE05} thus indicate a history of iron-rich dust production.



The 20-$\mu$m feature, seen in NGC 6352 V5, is also observed in NGC 5927 V1
and Terzan 5 V6. It has a width (FWHM) of $\sim 3 \mu$m and a
line-to-continuum ratio up to 50\%. This feature has previously been seen in
low mass-loss stars; it correlates with the presence of an 11-$\mu$m shoulder,
and 13.1- and 28-$\mu$m peaks \citep{LML88,SKGP03}. The related peaks are also
present in our sample, and we confirm that the 20-$\mu$m feature disappears
for more evolved (i.e.\ more luminous) stars. The 20-$\mu$m and associated features are barely
detectable in the naked stars, strongest in stars with weak silicate emission
and disappear when silicates dominate (Figure \ref{MbolFeHFig}). This suggests
they are produced either during the formation or destruction of silicate
grains. The identification of the feature with Mg$_x$Fe$_{1-x}$O is
unconfirmed mainly because it does not predict the associated features, but
the proposed identifications for those (corundum, pyroxenes) \citep{SKGP03} do
not provide good fits at 20 $\mu$m, making it plausible that a range of
minerals form under these conditions. We suggest that these features may be
caused by the chemical modification of silicate grains: silicate grains which adsorb
iron into the lattice have higher opacity, and can be heated until the iron-bearing component dissociates from the grain \citep{Woitke06b}. This is expected to occur at any radius from the star at which dust formation is efficient \citep{GS99}. In this model, metallic iron is produced first, but
at higher metallicity and mass loss rates, the iron becomes partly
incorporated into other grain types including iron-oxide and iron-containing
silicates.

Studies of inclusions within meteoric primitive Solar System material have
traced dust grains to specific origins in AGB stars, supernovae and novae,
through their isotopic ratios \citep{Zinner03,MKL05}. Recently, the first
pre-solar iron-oxide grain has been identified \citep{FSB08}. The iron and
oxygen isotopic ratios in it are indicative of an origin in an AGB star. This
supports the hypothesis that AGB stars are a source of interstellar iron dust,
and have contributed to the Solar System's iron content.

\subsection{The dust condensation sequence at low [Fe/H]}

\begin{figure}
\includegraphics[width=0.35\textwidth,angle=-90]{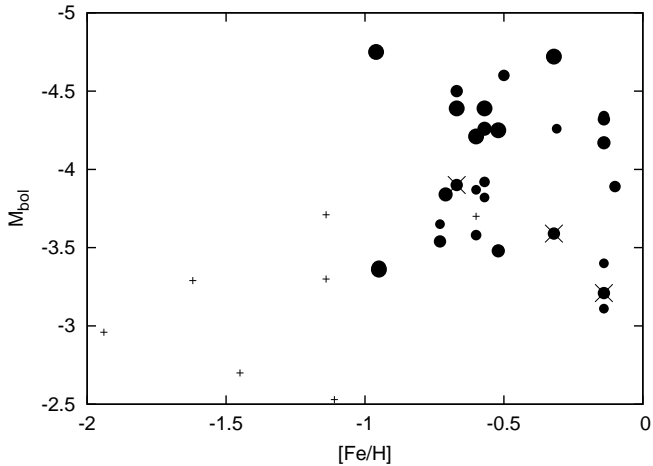}
\caption{Variation of dust properties with bolometric magnitude and metallicity. Symbols as Figure \ref{Excess2Fig}.}
\label{MbolFeHFig}
\end{figure}

Figure \ref{MbolFeHFig} shows the variation of dust types (see Paper I) with
bolometric magnitude and metallicity. Silicate-rich winds are only seen at
high luminosity or high metallicity, with iron dominating the outflow in
lower-luminosity, lower-metallicity stars. This indicates that the dust
condensation sequence depends on metallicity and luminosity. \citet{GS99}
show that, under equilibrium conditions, iron is only expected to condense before silicates in regions with high gas pressure. If this applies to our stars, it implies high gas pressure
at low metallicity and/or low luminosity, as may be expected if the outflow
velocities where iron condenses (at the outside edge of the molsphere) are
very low. Under non-equilibrium conditions, grain growth chemistry is strongly controlled by kinetics \citep{TNO+09}, which may favour the formation of iron.

In \citet{GS99}, silicates form before iron at lower densities
and temperatures. The restriction of silicates towards higher metallicity and
luminosity (Figure \ref{MbolFeHFig}) suggests that their circumstellar
envelopes are more extended, favouring silicates.  Silicate production is
expected to increase with both stellar metallicity and luminosity
\citep{vanLoon00}.


\section{Conclusions}

Using mid-infrared spectra and photometry, we have shown that there is a considerable amount of unattributed infrared flux in a large selection of globular cluster giants. Through radiative transfer modelling, we deduce that this flux is most likely due to metallic iron grains forming in a truncated stellar wind, corroborated by a potential identification of FeO with the 20-$\mu$m emission feature. The production of metallic iron seems to become more dominant at lower metallicity, suggesting that the dust condensation sequence is fundamentally different in metal-poor objects. Large-scale production of iron grains and iron oxide in AGB stars can explain iron depletion in the gas and solid phases of the post-AGB stars and planetary nebulae; as well as isotopic ratios in FeO grains in meteorites. While iron increases opacity in oxygen-rich winds, it remains unclear whether it can dominate driving of metal-poor winds. Future observations to determine outflow velocities in these stars should help determine the driver of these stellar winds.

\vspace{5 mm}

This paper uses observations made using the \emph{Spitzer Space Telescope}, operated by JPL, California Institute of Technology under NASA contract 1407 and supported by NASA through JPL (contract number 1257184). We thank Martha Boyer for her helpful comments.


\begin{thebibliography}{50}
\expandafter\ifx\csname natexlab\endcsname\relax\def\natexlab#1{#1}\fi

\bibitem[{{Beck} {et~al.}(1992){Beck}, {Gail}, {Henkel}, \&
  {Sedlmayr}}]{BGH+92}
{Beck}, H.~K.~B., {Gail}, H., {Henkel}, R., \& {Sedlmayr}, E. 1992, A\&A, 265,
  626

\bibitem[{{Begemann} {et~al.}(1997){Begemann}, {Dorschner}, {Henning},
  {Mutschke}, {Guertler}, {Koempe}, \& {Nass}}]{BDH+97}
{Begemann}, B., {Dorschner}, J., {Henning}, T., {Mutschke}, H., {Guertler}, J.,
  {Koempe}, C., \& {Nass}, R. 1997, ApJ, 476, 199

\bibitem[{{Bowen}(1988)}]{Bowen88}
{Bowen}, G.~H. 1988, in ASSL Vol. 148: Pulsation and Mass Loss in Stars, ed.
  R.~{Stalio} \& L.~A. {Willson}, 3

\bibitem[{{Bowen}(1989)}]{Bowen89}
{Bowen}, G.~H. 1989, in IAU Colloq. 106: Evolution of Peculiar Red Giant Stars,
  ed. {H.~R.~Johnson \& B.~Zuckerman}, 269--283

\bibitem[{{Boyer} {et~al.}(2008){Boyer}, {McDonald}, {van Loon}, {Woodward},
  {Gehrz}, {Evans}, \& {Dupree}}]{BMvL+08}
{Boyer}, M.~L., {McDonald}, I., {van Loon}, J.~T., {Woodward}, C.~E., {Gehrz},
  R.~D., {Evans}, A., \& {Dupree}, A.~K. 2008, AJ, 135, 1395

\bibitem[{{Boyer} {et~al.}(2009){Boyer}, {McDonald}, {van Loon}, {Gordon},
  {Babler}, {Block}, {Bracker}, {Engelbracht}, {Hora}, {Indebetouw}, {Meade},
  {Meixner}, {Misselt}, {Oliveira}, {Sewilo}, {Shiao}, \& {Whitney}}]{BMvL+09}
{Boyer}, M.~L., {et~al.} 2009, ApJ, 705, 746

\bibitem[{{Cami} {et~al.}(2010){Cami}, {van Malderen}, \& {Markwick}}]{CvMM10}
{Cami}, J., {van Malderen}, R., \& {Markwick}, A.~J. 2010, ApJS, 999, 999

\bibitem[{{Chlewicki} \& {Laureijs}(1988)}]{CL88}
{Chlewicki}, G., \& {Laureijs}, R.~J. 1988, A\&A, 207, L11+

\bibitem[{{Delgado Inglada} {et~al.}(2009){Delgado Inglada},
  {Rodr{\'{\i}}guez}, {Mampaso}, \& {Viironen}}]{DIRMV09}
{Delgado Inglada}, G., {Rodr{\'{\i}}guez}, M., {Mampaso}, A., \& {Viironen}, K.
  2009, ApJ, 694, 1335

\bibitem[{{Draine} \& {Lee}(1984)}]{DL84}
{Draine}, B.~T., \& {Lee}, H.~M. 1984, ApJ, 285, 89

\bibitem[{{Duari} {et~al.}(1999){Duari}, {Cherchneff}, \& {Willacy}}]{DCW99}
{Duari}, D., {Cherchneff}, I., \& {Willacy}, K. 1999, A\&A, 341, L47

\bibitem[{{Elitzur} {et~al.}(1989){Elitzur}, {Brown}, \& {Johnson}}]{EBJ89}
{Elitzur}, M., {Brown}, J.~A., \& {Johnson}, H.~R. 1989, ApJ, 341, L95

\bibitem[{{Flaherty} {et~al.}(2007){Flaherty}, {Pipher}, {Megeath}, {Winston},
  {Gutermuth}, {Muzerolle}, {Allen}, \& {Fazio}}]{FPM+07}
{Flaherty}, K.~M., {Pipher}, J.~L., {Megeath}, S.~T., {Winston}, E.~M.,
  {Gutermuth}, R.~A., {Muzerolle}, J., {Allen}, L.~E., \& {Fazio}, G.~G. 2007,
  ApJ, 663, 1069

\bibitem[{{Floss} {et~al.}(2008){Floss}, {Stadermann}, \& {Bose}}]{FSB08}
{Floss}, C., {Stadermann}, F.~J., \& {Bose}, M. 2008, ApJ, 672, 1266

\bibitem[{{Gail} \& {Sedlmayr}(1999)}]{GS99}
{Gail}, H., \& {Sedlmayr}, E. 1999, A\&A, 347, 594

\bibitem[{{Gustafsson} {et~al.}(1975){Gustafsson}, {Bell}, {Eriksson}, \&
  {Nordlund}}]{GBEN75}
{Gustafsson}, B., {Bell}, R.~A., {Eriksson}, K., \& {Nordlund}, A. 1975, A\&A,
  42, 407

\bibitem[{{Gustafsson} {et~al.}(2008){Gustafsson}, {Edvardsson}, {Eriksson},
  {J{\o}rgensen}, {Nordlund}, \& {Plez}}]{GEE+08}
{Gustafsson}, B., {Edvardsson}, B., {Eriksson}, K., {J{\o}rgensen}, U.~G.,
  {Nordlund}, {\AA}., \& {Plez}, B. 2008, A\&A, 486, 951

\bibitem[{Harris(1996)}]{Harris96}
Harris, W.~E. 1996, ApJ, 112, 1487

\bibitem[{{Henning} {et~al.}(1995){Henning}, {Begemann}, {Mutschke}, \&
  {Dorschner}}]{HBMD95}
{Henning}, T., {Begemann}, B., {Mutschke}, H., \& {Dorschner}, J. 1995, A\&AS,
  112, 143

\bibitem[{{H{\"o}fner}(2008)}]{Hoefner08}
{H{\"o}fner}, S. 2008, A\&A, 491, L1

\bibitem[{{H{\"o}fner} \& {Andersen}(2007)}]{HA07}
{H{\"o}fner}, S., \& {Andersen}, A.~C. 2007, A\&A, 465, L39

\bibitem[{{Kemper} {et~al.}(2002){Kemper}, {de Koter}, {Waters}, {Bouwman}, \&
  {Tielens}}]{KdKW+02}
{Kemper}, F., {de Koter}, A., {Waters}, L.~B.~F.~M., {Bouwman}, J., \&
  {Tielens}, A.~G.~G.~M. 2002, A\&A, 384, 585

\bibitem[{{Lebouteiller} {et~al.}(2008){Lebouteiller}, {Bernard-Salas},
  {Brandl}, {Whelan}, {Wu}, {Charmandaris}, {Devost}, \& {Houck}}]{LBSB+08}
{Lebouteiller}, V., {Bernard-Salas}, J., {Brandl}, B., {Whelan}, D.~G., {Wu},
  Y., {Charmandaris}, V., {Devost}, D., \& {Houck}, J.~R. 2008, ApJ, 680, 398

\bibitem[{{Lebzelter} {et~al.}(2006){Lebzelter}, {Posch}, {Hinkle}, {Wood}, \&
  {Bouwman}}]{LPH+06}
{Lebzelter}, T., {Posch}, T., {Hinkle}, K., {Wood}, P.~R., \& {Bouwman}, J.
  2006, ApJ, 653, L145

\bibitem[{{Little-Marenin} \& {Little}(1988)}]{LML88}
{Little-Marenin}, I.~R., \& {Little}, S.~J. 1988, ApJ, 333, 305

\bibitem[{{Maas} {et~al.}(2005){Maas}, {Van Winckel}, \& {Lloyd
  Evans}}]{MvWLE05}
{Maas}, T., {Van Winckel}, H., \& {Lloyd Evans}, T. 2005, A\&A, 429, 297

\bibitem[{{Mathis} {et~al.}(1977){Mathis}, {Rumpl}, \& {Nordsieck}}]{MRN77}
{Mathis}, J.~S., {Rumpl}, W., \& {Nordsieck}, K.~H. 1977, ApJ, 217, 425

\bibitem[{{McClure}(2009)}]{McClure09}
{McClure}, M. 2009, ApJ, 693, L81

\bibitem[{{McDonald} \& {van Loon}(2007)}]{MvL07}
{McDonald}, I., \& {van Loon}, J.~T. 2007, A\&A, 476, 1261

\bibitem[{{McDonald} {et~al.}(2009){McDonald}, {van Loon}, {Decin}, {Boyer},
  {Dupree}, {Evans}, {Gehrz}, \& {Woodward}}]{MvLD+09}
{McDonald}, I., {van Loon}, J.~T., {Decin}, L., {Boyer}, M.~L., {Dupree},
  A.~K., {Evans}, A., {Gehrz}, R.~D., \& {Woodward}, C.~E. 2009, MNRAS, 394,
  831

\bibitem[{{Messenger} {et~al.}(2005){Messenger}, {Keller}, \&
  {Lauretta}}]{MKL05}
{Messenger}, S., {Keller}, L.~P., \& {Lauretta}, D.~S. 2005, Science, 309, 737

\bibitem[{{Nenkova} {et~al.}(1999){Nenkova}, {Ivezi{\'c}}, \&
  {Elitzur}}]{NIE99}
{Nenkova}, M., {Ivezi{\'c}}, {\v{Z}}., \& {Elitzur}, M. 1999, LPI
  Contributions, 969, 20

\bibitem[{{Ordal} {et~al.}(1988){Ordal}, {Bell}, {Alexander}, {Newquist}, \&
  {Querry}}]{OBA+88}
{Ordal}, M.~A., {Bell}, R.~J., {Alexander}, Jr., R.~W., {Newquist}, L.~A., \&
  {Querry}, M.~R. 1988, Applied Optics, 27, 1203

\bibitem[{{Partridge} \& {Schwenke}(1997)}]{PS97}
{Partridge}, H., \& {Schwenke}, D.~W. 1997, J.~Chem.~Phys, 106, 4618

\bibitem[{{Pijpers} \& {Hearn}(1989)}]{PH89}
{Pijpers}, F.~P., \& {Hearn}, A.~G. 1989, A\&A, 209, 198

\bibitem[{{Posch} {et~al.}(2002){Posch}, {Kerschbaum}, {Mutschke}, {Dorschner},
  \& {J{\"a}ger}}]{PKM+02}
{Posch}, T., {Kerschbaum}, F., {Mutschke}, H., {Dorschner}, J., \& {J{\"a}ger},
  C. 2002, A\&A, 393, L7

\bibitem[{{Reid} \& {Menten}(1997)}]{RM97}
{Reid}, M.~J., \& {Menten}, K.~M. 1997, ApJ, 476, 327

\bibitem[{{Savage} \& {Bohlin}(1979)}]{SB79}
{Savage}, B.~D., \& {Bohlin}, R.~C. 1979, ApJ, 229, 136

\bibitem[{{Sloan} {et~al.}(2003){Sloan}, {Kraemer}, {Goebel}, \&
  {Price}}]{SKGP03}
{Sloan}, G.~C., {Kraemer}, K.~E., {Goebel}, J.~H., \& {Price}, S.~D. 2003, ApJ,
  594, 483

\bibitem[{{Sloan} {et~al.}(2010){Sloan}, {Matsunaga}, {Matsuura}, {Zijlstra},
  {Kraemer}, {Wood}, {Nieusma}, {Bernard-Salas}, {Devost}, \& {Houck}}]{SMM+10}
{Sloan}, G.~C., {et~al.} 2010, ApJ, 999, 9999

\bibitem[{{Smith} {et~al.}(1999){Smith}, {Shetrone}, \& {Keane}}]{SSK99}
{Smith}, V.~V., {Shetrone}, M.~D., \& {Keane}, M.~J. 1999, ApJ, 516, L73

\bibitem[{{Tachibana} {et~al.}(2009){Tachibana}, {Nagahara}, {Ozawa}, {Tamada},
  \& {Ogawa}}]{TNO+09}
{Tachibana}, S., {Nagahara}, H., {Ozawa}, K., {Tamada}, S., \& {Ogawa}, R.
  2009, in Lunar and Planetary Inst. Technical Report, Vol.~40, Lunar and
  Planetary Institute Science Conference Abstracts, 2512--+

\bibitem[{{Tsuji}(2000)}]{Tsuji00}
{Tsuji}, T. 2000, ApJ, 540, L99

\bibitem[{{Tsuji} {et~al.}(1997){Tsuji}, {Ohnaka}, {Aoki}, \&
  {Yamamura}}]{TOAY97}
{Tsuji}, T., {Ohnaka}, K., {Aoki}, W., \& {Yamamura}, I. 1997, A\&A, 320, L1

\bibitem[{{van Loon}(2000)}]{vanLoon00}
{van Loon}, J.~T. 2000, A\&A, 354, 125

\bibitem[{{Verhoelst} {et~al.}(2009){Verhoelst}, {van der Zypen}, {Hony},
  {Decin}, {Cami}, \& {Eriksson}}]{VvdZH+09}
{Verhoelst}, T., {van der Zypen}, N., {Hony}, S., {Decin}, L., {Cami}, J., \&
  {Eriksson}, K. 2009, A\&A, 498, 127

\bibitem[{{Voshchinnikov} \& {Henning}(2008)}]{VH08}
{Voshchinnikov}, N.~V., \& {Henning}, T. 2008, A\&A, 483, L9

\bibitem[{{Whittet}(1992)}]{Whittet92}
{Whittet}, D.~C.~B. 1992, {Dust in the galactic environment} (Institute of
  Physics Publishing.)

\bibitem[{{Woitke}(2006)}]{Woitke06b}
{Woitke}, P. 2006, A\&A, 460, L9

\bibitem[{{Zinner}(2003)}]{Zinner03}
{Zinner}, E.~K. 2003, Treatise on Geochemistry, 1, 17

\end{thebibliography}

\end{document}